\documentclass{elsart}
\usepackage{graphicx}
\usepackage{times}

\def\la{\mathrel{\mathpalette\fun <}}
\def\ga{\mathrel{\mathpalette\fun >}}
\def\fun#1#2{\lower3.6pt\vbox{\baselineskip0pt\lineskip.9pt                                           
\ialign{$\mathsurround=0pt#1\hfil##\hfil$\crcr#2\crcr\sim\crcr}}}

\newcommand{\bc}{\begin{center}}
\newcommand{\ec}{\end{center}}
\newcommand{\bd}{\begin{displaymath}}
\newcommand{\ed}{\end{displaymath}}
\newcommand{\be}{\begin{equation}}
\newcommand{\ee}{\end{equation}}
\newcommand{\ba}{\begin{array}}
\newcommand{\ea}{\end{array}}
\newcommand{\bt}{\begin{tabular}}
\newcommand{\et}{\end{tabular}}

\begin{document}

\begin{frontmatter}
\title{Inclusive pion double charge exchange on $^{\mathbf{16}}$O\\ 
at 0.6 $-$ 1.1 GeV}
\author{B.M. Abramov},
\author{Yu.A. Borodin}, 
\author{S.A. Bulychjov}, 
\author{I.A. Dukhovskoy}, 
\author{A.B. Kaidalov},
\author{A.P. Krutenkova}, 
\author{V.V. Kulikov}, 
\author{M.A. Matsyuk}, 
\author{I.A. Radkevich}\footnote[1]{Deceased}, 
\author{E.N. Turdakina}
\address{State Research Center -- Institute for Theoretical and
Experimental Physics,\\ 117218 Moscow, Russian Federation}
\author{L. Alvarez-Ruso}
\address{Nuclear Science Division,  Lawrence Berkeley National Laboratory
\\ 1 Cyclotron Rd., Berkeley, CA 94720, USA}
\author{M.J. Vicente Vacas}
\address{Departamento de F\'{\i}sica Te\'orica and IFIC, Centro Mixto
Univesidad de Valencia-CSIC,\\ Ap. Correos 22085, 46071 Valencia, Spain}


\begin{abstract}
The inclusive pion double charge exchange (DCX) on oxygen nuclei
has been measured 
in the region where additional pion production is kinematically forbidden.
The experiment was performed at the ITEP PS at incident $\pi^-$  kinetic energies 
$T_0=$ 0.59, 0.75 and 1.1 GeV. The integrated forward differential cross section 
was found to decrease with energy slowly. At 1.1 GeV it exceeds the theoretical 
prediction  within the conventional sequential single charge exchange mechanism 
with a neutral pion in the intermediate state (Glauber elastic
rescattering)  
by about a factor of five. 
The sequential mechanism with two pions in 
the intermediate state  (Glauber inelastic rescatterings), which was proposed 
recently, seems to be able to explain the observed energy dependence 
and allows to 
predict the DCX cross section at higher energies.
\end{abstract}
\end{frontmatter}

\section{Introduction}
Although pion double charge exchange (DCX) reactions on nuclei
have long been studied,
 understanding the reaction mechanism within  models of pion propagation through a
nucleus appears to be a complicated problem because of the two-nucleon nature of the process.
Nevertheless the unique feature of pion DCX  of
requiring at least two like nucleons of a nucleus to participate 
provides a good testing ground for an  investigation of different few-body
and nuclear structure effects
in the nucleus (see \cite{jibuti}-\cite{johnson} and references
therein).
The pion DCX process
was first suggested as a probe of short range nucleon-nucleon correlations by De Shalit,
Drell and Lipkin in 1961 \cite{deshali}. The pioneering measurement of pion DCX was done 
at JINR \cite{batus} in 1963 at 30 -- 80 MeV in inclusive
reactions. The next step was made 
at meson factories which gave start to the investigation of exclusive DCX reactions since 1977.
While these reactions were mainly used for the study of nuclear structure effects (excitation 
of double isobar-analog transitions, search for exotic nuclear states etc.), both inclusive 
and exclusive DCX measurements are important to study hadron dynamics. 

For energies up to 0.5 GeV the conventional mechanism of two sequential single charge exchanges 
(SSCX) has traditionally been able
to explain the main features of pion DCX. The thorough 
experimental investigations of the differential cross sections of inclusive DCX reactions, 
$A(\pi^{\pm},\pi^{\mp})\mathrm{X}$, for a wide range of nuclei from $^3$He to $^{208}$Pb in 
the energy region of $\Delta$ resonance (see Ref. \cite{gram}-\cite{yuly} and the recent 
theoretical analysis \cite{gibbs}) confirmed that the SSCX 
is the dominant mechanism. The rather smooth energy dependence of forward exclusive DCX cross 
section from 0.3 to 0.5 GeV ~\cite{johnson}, \cite{williams} was successfully reproduced by 
Oset et al. in \cite{oset} within SSCX calculations in the Glauber approach using known 
$\pi \mathrm{N}$ amplitudes. For higher energies these authors  \cite{oset}
predicted a strong 
decrease
of forward exclusive DCX cross section. Any significant deviation from the expected low value 
of the DCX cross section near 1 GeV offers a unique possibility to search for  
nonconventional DCX mechanisms (short range NN correlations, in particular) \cite{oset}, \cite{hashimoto}.

We present here an experimental study of
inclusive pion DCX at energies above 0.5~GeV, where there are no data
available.  Inclusive processes have much larger cross sections than
exclusive ones and their registration does not require an excellent energy resolution. So, they
are within existing experimental possibilities. Theoretically, the energy behaviour
of the inclusive reactions seems less sensitive to the details of the nuclear structure and
within the SSCX model
reflects the energy dependence of the $\pi \mathrm{N}$ single charge exchange (SCX)
amplitude \cite{vacas}. At high energy, the DCX reactions 
can be studied in the kinematical region near the
high-energy end point of the pion spectra, where pion
production (the $(\pi,\pi\pi)$ process) on one nucleon is
kinematically forbidden. 
The few
existing experiments on DCX at high energies \cite{leksin,gean} have no events
in this specific region of true DCX. 
For example, the events measured at
$T_\pi = 1.7$~GeV on $^4$He~\cite{gean} belong to the region where both true
DCX and two-pion production on a single nucleon coexist. Actually, the
analysis performed in Ref.~\cite{gean} reveals the dominance of the later
mechanism  followed by a two-body pion absorption at this energy.

The goal of this study is to get an information to
clarify the mechanism of pion DCX on nuclei in the energy region near 1~GeV. Our 
preliminary results \cite{fbsuppl} and \cite{Yaf}, 
obtained at the ITEP PS definitely showed that at 0.75 GeV and at 1.1 GeV we observed 
a pion DCX signal at a higher level than predicted in the framework of the SSCX model. The attempts 
\cite{luis} and \cite{kaid} to explain this observation have led to the suggestion of a new 
mechanism \cite{kaid} of pion DCX at energies above 0.6 GeV.

The results considered below are based on our full statistics of $\pi^+$ energy 
spectra from the reaction
\begin{equation}
\pi^- +{}^{16}\mathrm{O} \rightarrow \pi^{+} + \mathrm{X} 
\end{equation}      
at $T_0=$ 0.59, 0.75 and 1.1 GeV (scattering angle $\theta=$ 0 -- 14$^o$). 

The paper is organized as follows. The experimental setup and data taking procedure are 
described in Section 2. Event selection is presented in Section 3. The DCX cross section 
calculation and an analysis of our results on $\pi^+$ spectra are given in Section 4.
In Section 5 the comparison of the experimental data with the 
SSCX model is made. 
Energy dependence of DCX cross section 
is discussed in Section 6. Nonconventional mechanisms of DCX and the theoretical interpretation 
of the results are discussed in Section 7. Conclusion is given in Section 8.

\section{Experimental setup and data taking}

\begin{figure}[bhtp]
\begin{center}
\includegraphics[bb=0 0 567 841,width=\textwidth]{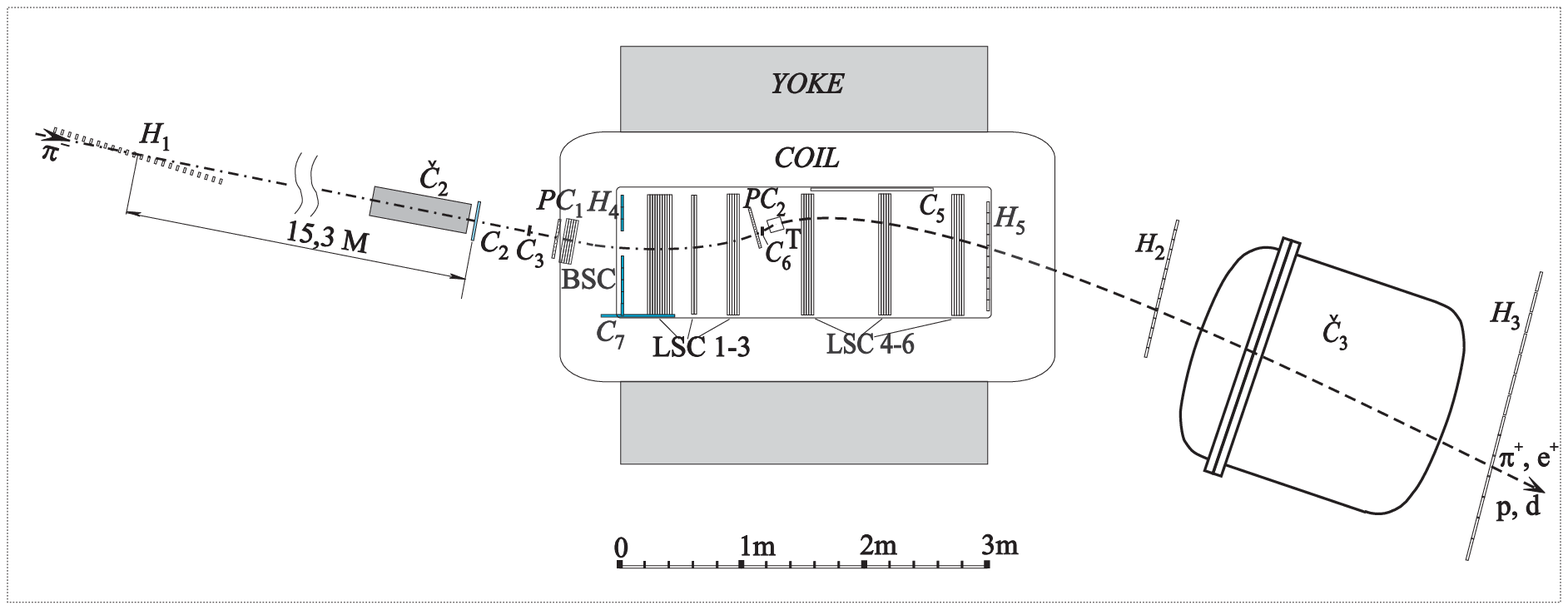}
\vspace*{-3.0cm}
\caption [] {The layout of the experimental setup.}
\label{diagram1}
\end{center}
\end{figure}

The data were taken at  the ITEP 10-GeV Proton Synchrotron using
negative pion beams with momenta 
$p_0=$ 0.72, 0.88 and 1.26 GeV/$c$ ($T_0=$ 0.59, 0.75 and 1.1 GeV).\footnote{The values 
of $p_0$ and $T_0$ are related to the middle of the target.} The beam flux was 
(1 - 5) $\times$10$^5$ pions per 1 second spill. The experiment was done at the
3m magnet spectrometer instrumented with multigap optical spark chambers \cite{mts}. 
The layout of the experimental setup is shown in Fig.~\ref{diagram1}.
The beam pions,  
defined by scintillation hodoscope $H_1$ and scintillation counters
$C_2$ and $C_3$, 
struck the target placed in the middle of the magnet. The hodoscope  $H_1$ (of 24 
scintillation counters placed at the intermediate focus of the beam) made it possible to 
determine a beam pion momentum with a precision of $\pm$0.3\%, while the full momentum 
acceptance of the beam was $\mathrm{\Delta}p/p \sim$5\%. The scintillation counter $C_6$ 
was used only for the beam adjustment. During the runs, the proportional chambers $PC_1$ and 
$PC_2$ served for the on-line monitoring of the beam position at the entrance of the
target. Most of the beam particles passing through the target without
interaction were vetoed 
by the counter $C_5$. $\mathrm{\breve{C}}$erenkov counter $\breve{C}_2$ 
filled with freon-12 was used to 
reject electrons. The electron contamination in the beam was 10\%, 6.5\% and 3\% at 
$p_0=$ 0.72, 0.88 and 1.26 GeV/$c$, respectively. Beam particle trajectories were reconstructed
with the help of the beam spark chamber (BSC, 10 gaps, 1 gap = 0.8 cm) and 
large spark chambers (LSC 1 -  3, 32 gaps in total) placed in the magnetic field. 

The  H$_2$O and D$_2$O targets were contained in the identical cylindrical cells (with 
8 cm in diameter and 9.5 cm in length) made of 0.014 cm stainless steel. They were positioned 
on the turnplate together with  $^6$Li, $^7$Li, $^{12}$C and ``empty'' targets. During 
each run they were substituted one for another at regular time intervals. The preliminary 
results with  $^6$Li and $^7$Li targets can be found in  \cite{fbsuppl} and \cite{litii}. 
The results for $^{12}$C will be published elsewhere.                           

Positive particles from the reactions 
\begin{equation}
\pi^- +  A \rightarrow \mathrm{(e^+}, \pi^+, \mathrm{p, d}) +  \mathrm{X}, 
\end{equation}  
emitted from the target in the forward direction 
were detected in 
three planes of scintillation hodoscopes $H_5$, $H_2$ and $H_3$ (14 
elements with overall area $\sim$1.5 m$^2$).
A time-of-flight (TOF) between $C_2$ and 
$H_3$ (on the base of $\sim$6 m)
was measured to select pions, protons and deuterons
with the appropriate triggers\footnote{The proton and the deuteron triggers used the 
scintillation hodoscope $H_{4}$ and the scintillation counter $C_{7}$ in coincidence
to detect backward $\pi^{-}$. With these triggers backward quasielastic 
$\pi^-$p and $\pi^-$d scattering
was studied. The preliminary results were published in \cite{quasi} 
and in \cite{quasipid}, respectively.}. The outgoing particle momentum was measured 
in the large spark chambers LSC 4 - 6 (30 gaps in total), placed in the magnet. 
For each beam momentum the current in the magnet was set proportional to the beam 
momentum to have the same space topology of the events in the spectrometer.
Pions were discriminated from protons/deuterons using the TOF measurement and from 
positrons using signals from the high pressure threshold 
$\mathrm{\breve{C}}$erenkov counter  
$\breve{C}_3$ \cite{cheren} 
of 1.6 m in diameter filled with freon-12.
 
Sparks, coded hodoscope hits and a TOF code (and some
other information) were photographed by a fast camera. All pictures without preliminary 
scanning were measured 
by a flying-spot digitizer PSP-2. Spark positions  were reconstructed in space and 
the track parameters were calculated. Event recognition, reconstruction 
and DST (data summary tape) production were carried out with the set of computer codes developed 
for our spectrometer \cite{recon}.
For each event the information written to the DST included
the momenta of incoming and outgoing particles at the vertex, vertex coordinates, hodoscope 
hits, TOF values, etc.  The analysis of the DST was carried out with the help of
the HBOOK package.

About 37000 pictures were taken with the H$_2$O and D$_2$O targets with pion
triggers
\begin{equation} 
S_{\pi} = (H_1C_2C_3\bar{\breve{C}_2})\bar C_5
(H_5H_2H_3\bar{\breve{C}_3})(C_{2}H_{3})_{t\pi},
\end{equation}
where $\mathrm{\breve{C}}$erenkov counters $\breve C_2$ and  $\breve C_3$ were used to reject electrons
/positrons, and
\begin{equation} 
S'_{\pi} = (H_1C_2C_3)\bar C_5(H_5H_2H_3)(C_{2}H_{3})_{t\pi};
\end{equation}              
both triggers selected pions by the TOF coincidence window $(C_2H_3)_{t\pi}$. 
These triggers worked simultaneously with the proton and the deuteron 
triggers (see footnote 2). With the $S_{\pi}$ trigger 
not only the 
cross section of the reaction (1) was measured but also the positron background 
(using the tagged $\breve C_3$ signal). The measured positron background
allowed to extract the cross section of the reaction (1) also from the events of
the $S'_{\pi}$ trigger.
However $\mathrm{\breve{C}}$erenkov counters limited the acceptance of the setup for
the proton and the deuteron triggers. So, a part of the statistics
was taken with the $S_{\pi}$ trigger and another part with the $S'_{\pi}$ trigger. 

The numbers of events,  $N_{\mathrm{DST}}$,
written to the DST with the $S_\pi$ trigger were 22060, 
2456 and 1752 at $T_0$ = 0.59 , 0.75 and 1.1  GeV, respectively, and with the
$S'_\pi$ trigger were 5395 and 5073 at $T_0$ = 0.75 and 1.1 GeV.


\section{$\pi^{-}\,\mathrm{{^{16}}O} \rightarrow \pi^{+}\,\mathrm{X}$ event selection}

Events of the reaction (1), which has relatively low cross section,
constituted a small part of 
$N_{\mathrm{DST}}$,
the main part came from the beam scattering 
off the magnet coil and yoke. 
The contribution of these events, as well 
as of those originating from the outgoing particle interaction in the 
material of hodoscopes $H_5$, $H_2$ and $H_3$, was 
effectively suppressed 
by two requirements: the trajectory of forward going particle should be 
successfully reconstructed in all three spark chambers LSC 4 - 6, and 
the  extrapolated trajectory should cross the hit elements of the hodoscopes.
The final sample contained the events of the reaction (1), proton background
from the reaction (2) and also positron background (only for the $S_\pi '$ trigger).

\begin{figure}[bhtp]
\begin{center}
\includegraphics[bb=0 0 567 841,width=\textwidth]{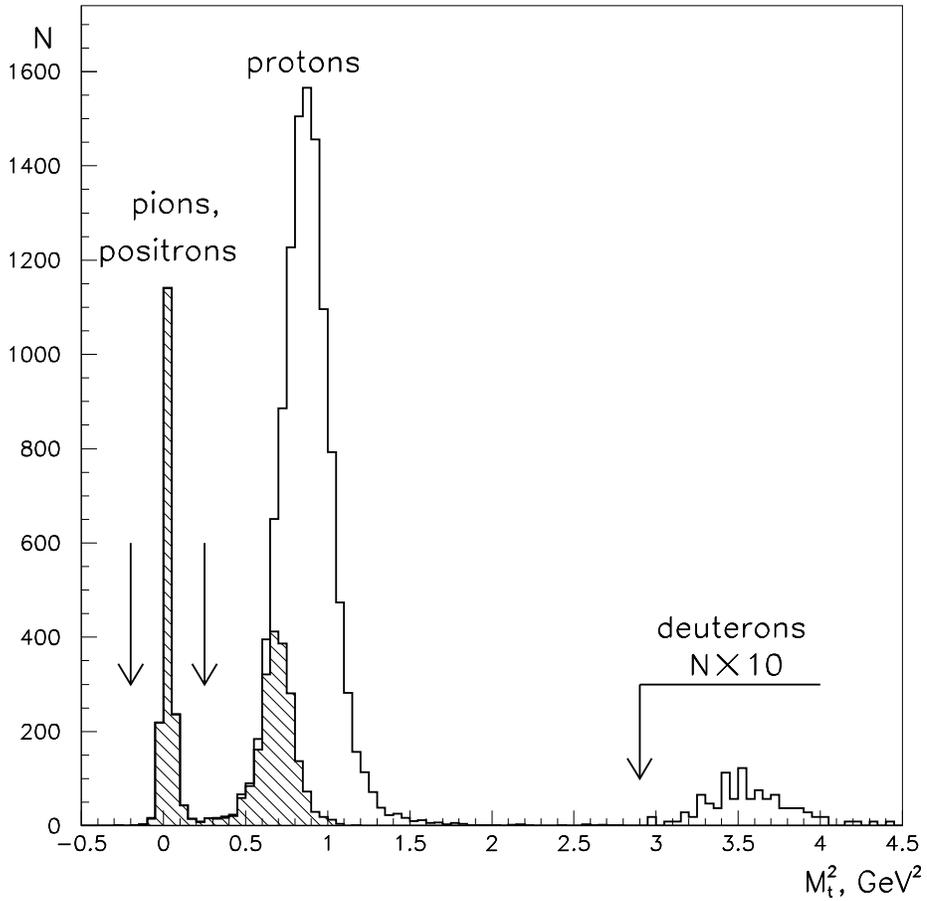}
\vspace*{-3.0cm}
\caption [] {The mass squared of positive particle{\bf s} from the reaction
$\pi^- A \rightarrow (\mathrm{e}^+, \pi^+, \mathrm{p, d) X}$, 
for the sum of events taken with a pion, 
a proton and a deuteron
triggers at $T_0$ = 0.59 GeV. The hatched histogram corresponds
to the pion ($S_{\pi}$) trigger.}
\label{diagram2}
\end{center}
\end{figure}

\begin{figure}[bhtp]
\begin{center}
\includegraphics[bb=0 0 567 841,width=\textwidth]{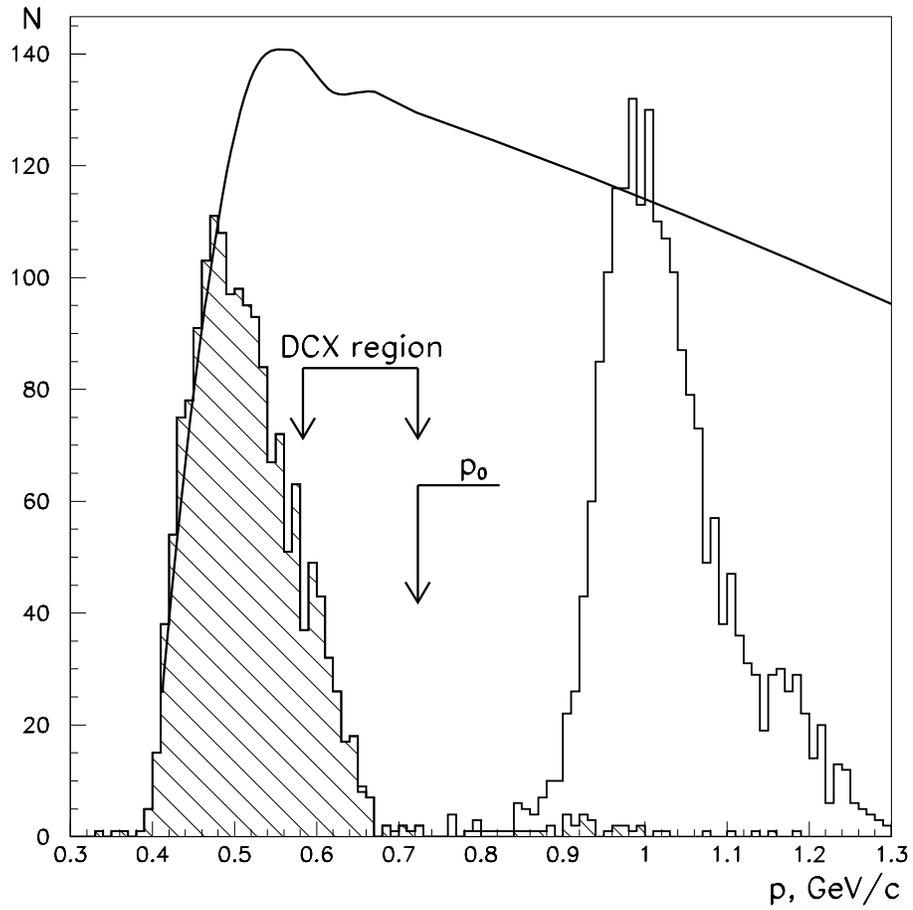}
\vspace*{-3.0cm}
\caption [] {The momentum distributions of the outgoing particle
in the $S_{\pi}$ trigger at $T_0$ = 0.59 GeV.
The hatched histogram is after the cut on $M_{t}^{2}$ to reject protons.
 The solid curve is the acceptance
of the apparatus (arbitrary units).}
\label{diagram3}
\end{center}
\end{figure}

The proton background was mainly suppressed by the TOF coincidence 
$(C_2H_3)_{t\pi}$ on the trigger level. The typical distribution of $M_t^2$, 
the outgoing particle mass squared\footnote{$M_t$ was calculated using the relation  
$M_t^2$ = $p^2(1-\beta^2)/\beta^2$, where $p$ is a particle momentum measured 
by the spectrometer and $\beta$ is the velocity determined by the TOF measurement
(1/$\beta$ = $\alpha(N-N_0$), where $N$ is the TDC channel number and $\alpha$ and 
$N_0$ are known constants).}, is shown in Fig.~\ref{diagram2} for a sum of pion 
($S_{\pi}$), proton and deuteron triggers taken at $T_0=$ 0.59 GeV. The hatched 
area in Fig.~\ref{diagram2} shows the events of the pion trigger, which accepted 
also protons with relatively high momenta. In the momentum distribution of these 
events (see Fig.~\ref{diagram3}) the two groups of events within the momentum 
acceptance of the apparatus (solid curve) are protons (at higher momenta) and 
pions. The cut on $M_t^2$ (events between the arrows in Fig.~\ref{diagram2}) 
practically suppressed the events to the right of the beam momentum  (protons)
(see hatched histogram in Fig.~\ref{diagram3}). The remaining events in this 
region were considered as the proton background from the non Gaussian TOF tail.   

The DCX candidates for the reaction (1) lie at the high-energy part of the 
pion spectrum in the kinematical region $\Delta T = T_0 - T \le m_{\pi} \simeq 140$ MeV 
($T$ is the kinetic energy of outgoing pion), where the additional pion 
production is forbidden. The number of the DCX candidates
was 434, 77 and 15 for  $T_0$ =  0.59, 0.75 and 1.1 GeV. The 
proton background
in the DCX region was estimated by the extrapolation of the distribution with 
the momentum higher than $p_0$ to the lower momentum region (for  $T_0$ =  0.59 
GeV see hatched histogram in  Fig.~\ref{diagram3}). The extrapolation function 
was taken from the momentum distribution of protons from the reaction 
$\pi^-$p$\rightarrow$pX measured separately. This function  weakly decreased with 
the decrease of momentum. The contribution of the proton background events to the DCX region
increased with the beam momentum and appeared to be (6 $\pm$ 1)\%, (13 $\pm$ 
4)\% and (20 $\pm$ 9)\% for $T_0$ =  0.59, 0.75 and 1.1 GeV in the $S_{\pi}$ 
trigger. The positron background was rejected in this trigger with the use of 
$\mathrm{\breve{C}}$erenkov counters.

\begin{figure}[bhtp]
\begin{center}
\includegraphics[bb=0 0 567 841,width=\textwidth]{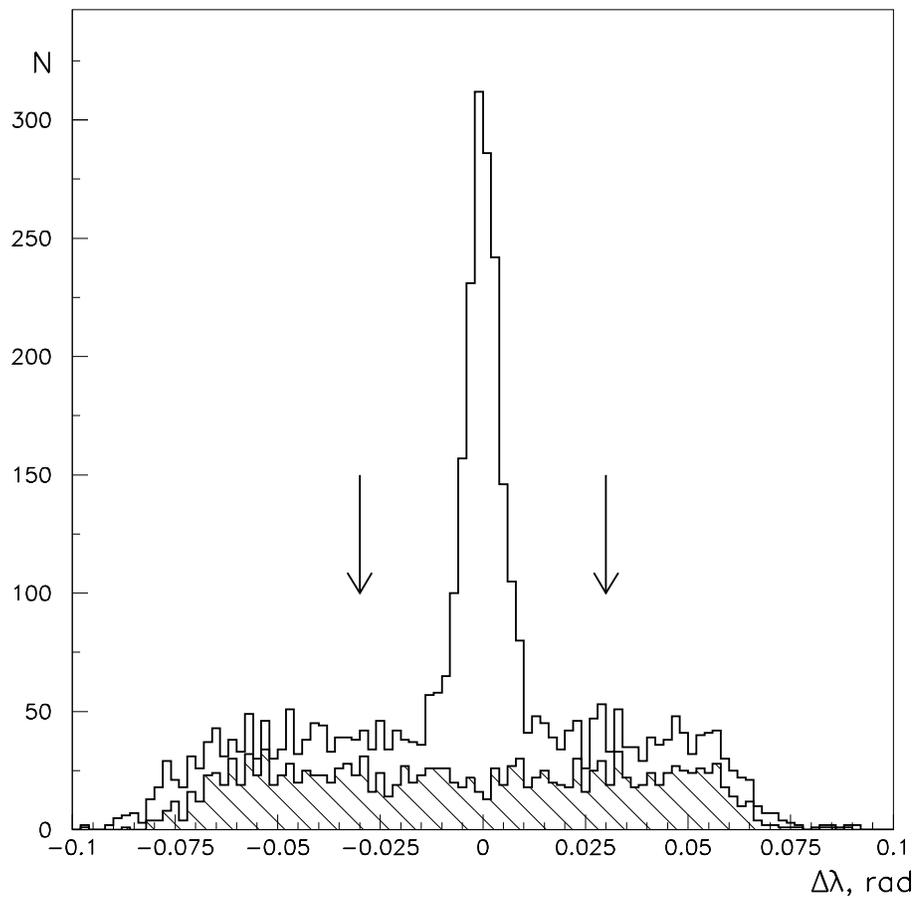}
\vspace*{-3.0cm}
\caption [] {The 
vertical projection
of the reaction angle of the outgoing particles (e$^+, \pi^+$ and p)
in the $S'_{\pi}$ trigger. The hatched histogram is for the protons.
Vertical arrows correspond to the $\Delta \lambda$ value of $\pm$ 0.03 rad.}
\label{diagram4}
\end{center}
\end{figure}

In order to select the events of the reaction (1) in the S$_{\pi}'$ trigger, it 
was necessary also to eliminate the beam positron background, BPB, and to  
subtract the target positron background, TPB. The BPB came from the electron 
admixture  in the beam, which produced fast forward positrons via bremsstrahlung 
in the target (following by the $\gamma$ conversion into 
$\mathrm{e^+_{forward} e^-}$ pair).
Forward positrons gave a sharp peak in the beam direction while pions and protons 
had a flat distribution within the acceptance. In Fig.~\ref{diagram4} we 
show the distributions of outgoing positive particles (pions, protons and 
positrons) on the vertical projection of the reaction angle, $\Delta \lambda$
(contrary to the horizontal one it was not distorted by the influence 
of the magnetic field). The hatched histogram in this figure 
stands for protons. The width of the zero-angle positron peak was
determined by multiple scattering in the target ($\sigma_{\Delta \lambda} \la$ 0.005 rad for water 
target at $T_0=$ 1.1 GeV). The cut $|\Delta \lambda| \ge$ 0.03 rad (see vertical arrows in Fig.~\ref{diagram4}) 
rejected the BPB in the $S_\pi '$ trigger to a
level of less than 3\%.

\begin{figure}[bhtp]
\begin{center}
\includegraphics[bb=0 0 567 841,width=\textwidth]{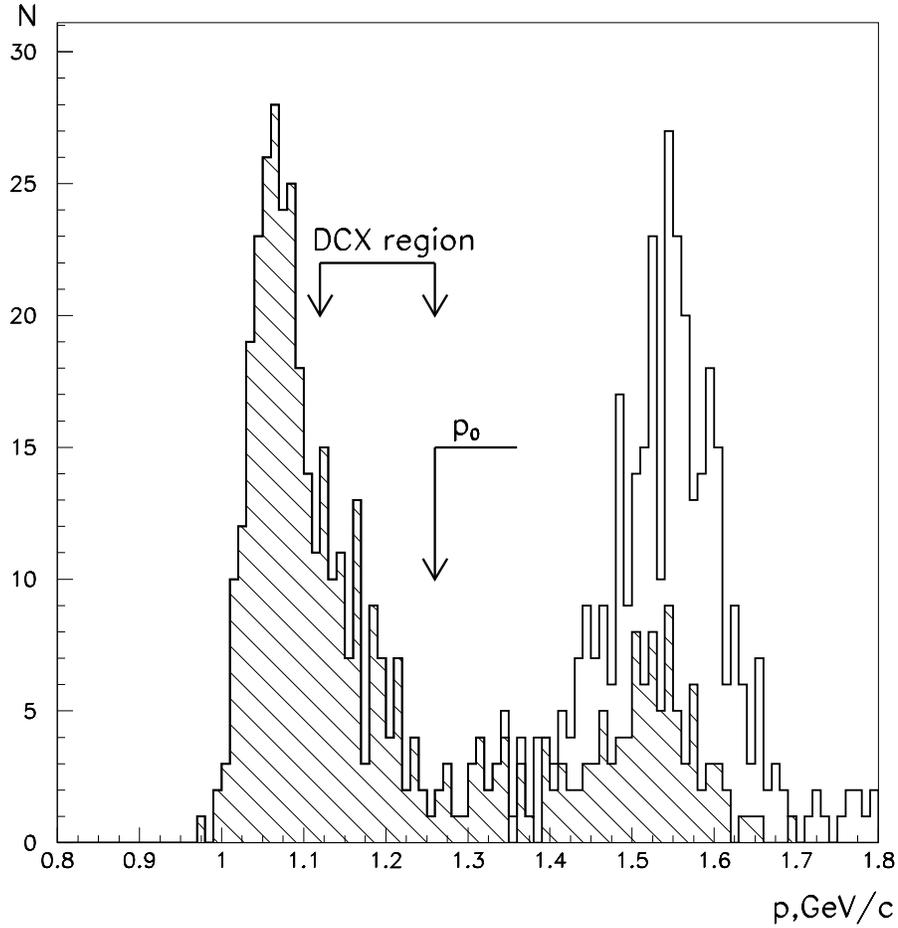}
\vspace*{-3.0cm}
\caption [] {The momentum distributions of the outgoing particles
in the $S'_{\pi}$ trigger at $T_0$ = 1.1 GeV after the cut 
$|\Delta \lambda| \ge $ 0.03 rad 
to reject beam positron background.
The hatched histogram is after the cut on $M_t^2$
to reject protons. $p_0$ is the beam momentum.}
\label{diagram5}
\end{center}
\end{figure}

The momentum distribution, analogous to the one presented in Fig.~\ref{diagram3} 
for  $T_0=$ 0.59 GeV and the $S_{\pi}$ trigger is shown in Fig.~\ref{diagram5} 
for  $T_0=$ 1.1 GeV and the $S_{\pi}'$ trigger events. 
Here, the additional cut $|\Delta \lambda| \ge$ 0.03 rad was applied to reject BPB
(for more details see \cite{acta}). 
The DCX region in this case contains events 
of the reaction (1) and of proton and positron (TPB) backgrounds. The number 
of the DCX candidates
was 
45 and 57 respectively for  
$T_0$ =  
0.75 and 1.1 GeV. The contribution of the proton background events to 
the DCX region was 
(10 $\pm$ 5)\% and (15 $\pm$ 7)\% for 
$T_0$ =  
0.75 and 1.1 GeV, respectively.

TPB is induced by the beam  pion interaction in the target
($\pi^-A$ $\rightarrow \pi^0A'$, 
$\pi^0 \rightarrow \gamma \gamma, \gamma \rightarrow \mathrm{e^+_{forward}e^-}$ 
or from Dalitz decay $\pi^0 \rightarrow \mathrm{e^+_{forward}e^-}\gamma$).
The use of  $\mathrm{\breve{C}}$erenkov detector $\breve C_3$ with the $S_{\pi}$ trigger suppressed 
this background, while in the $S'_{\pi}$ trigger we should take it into account.
To estimate TPB we analyzed the $\breve C_3$ signal in the $S_{\pi}$ trigger 
in order to select events with forward going positrons and pions. The 
$\breve C_3$ signals due to knockout of $\delta$ electrons
in $\breve C_3$ 
and to the noise of its PMTs were also taken into account. 
The ratio
of TPB events to the DCX candidates (pions plus TPB) for $|\Delta \lambda| \ge $ 0.03 rad 
was 
(12 $\pm$ 4)\% and (14 $\pm$ 8)\% for   
$T_0 $ = 0.75 and 1.1 GeV, respectively.  
\section{Cross section calculation}

\begin{figure}[bhtp]
\begin{center}
\includegraphics[bb=0 0 567 841,width=\textwidth]{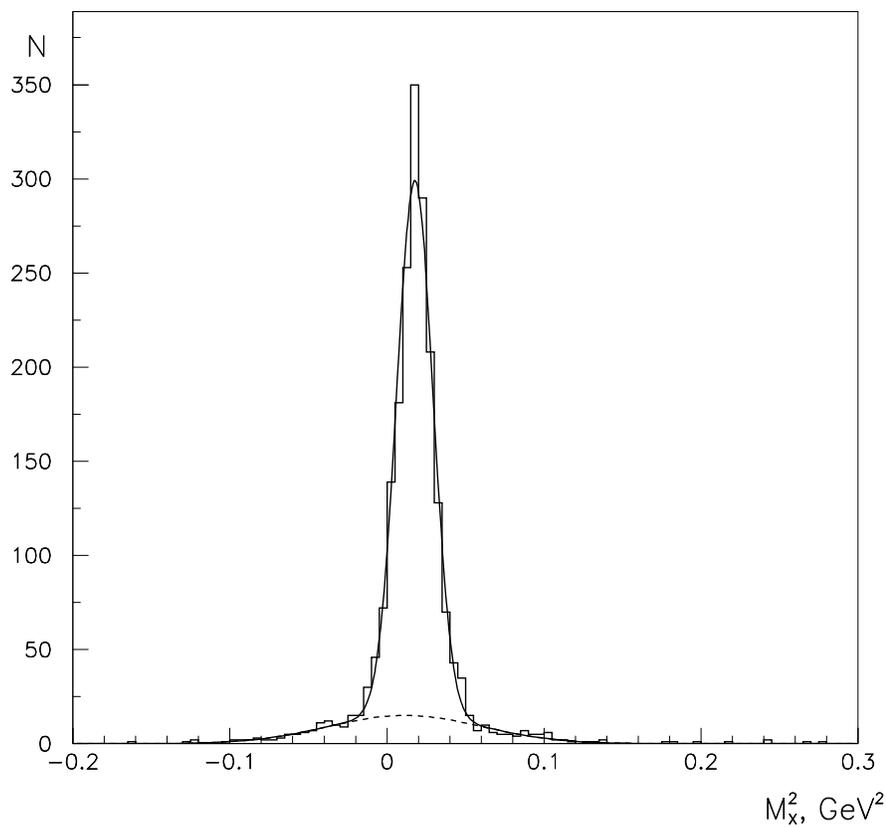}
\vspace*{-2.0cm}
\caption [] {Missing mass squared  for the reaction $\pi^- \mathrm{p} \rightarrow
 \mathrm{p}$X on the H$_2$O target. 
The solid curve is a two-Gaussian fit with $\sigma$ = 0.011 and
0.049, and peak positions of
0.018 and 0.011 GeV$^2$ 
respectively. The narrower Gaussian describes the elastic (pion) peak
while the other, shown by the dashed curve, stands for the oxygen background.}
\label{diagram6}
\end{center}
\end{figure}

The events which satisfied the requirements of the previous Section were taken
as candidates for the reaction (1). In
order to facilitate the comparison
of the data at different energies, we used event distributions on
 $\Delta T = T_0 - T$ ($\Delta T > 0$). The calibration of the $\Delta T$
scale was checked using the position of the backward elastic $\pi^-$p
scattering peak  in the reaction $\pi^- \mathrm{p} \rightarrow \mathrm{p}$X
(M$^2_{\mathrm{X}}$ = m$_{\pi}^2 \simeq$ 0.0195 GeV$^2$)
for the proton trigger  on the water target. The $\Delta T$ resolution
was estimated from the width of this peak, and
varied from 6 to 8 MeV for $T_0 $ from 0.59 to 1.1 GeV. As an example,
in Fig.~\ref{diagram6} the proton spectrum obtained at $T_{0}$
 = 0.59 GeV was 
fitted with a sum of two Gaussians (solid curve), 
one for the proton peak and another for the oxygen background (dashed curve). 

\begin{figure}[bhtp]
\begin{center}
\includegraphics[bb=0 0 567 841,width=\textwidth]{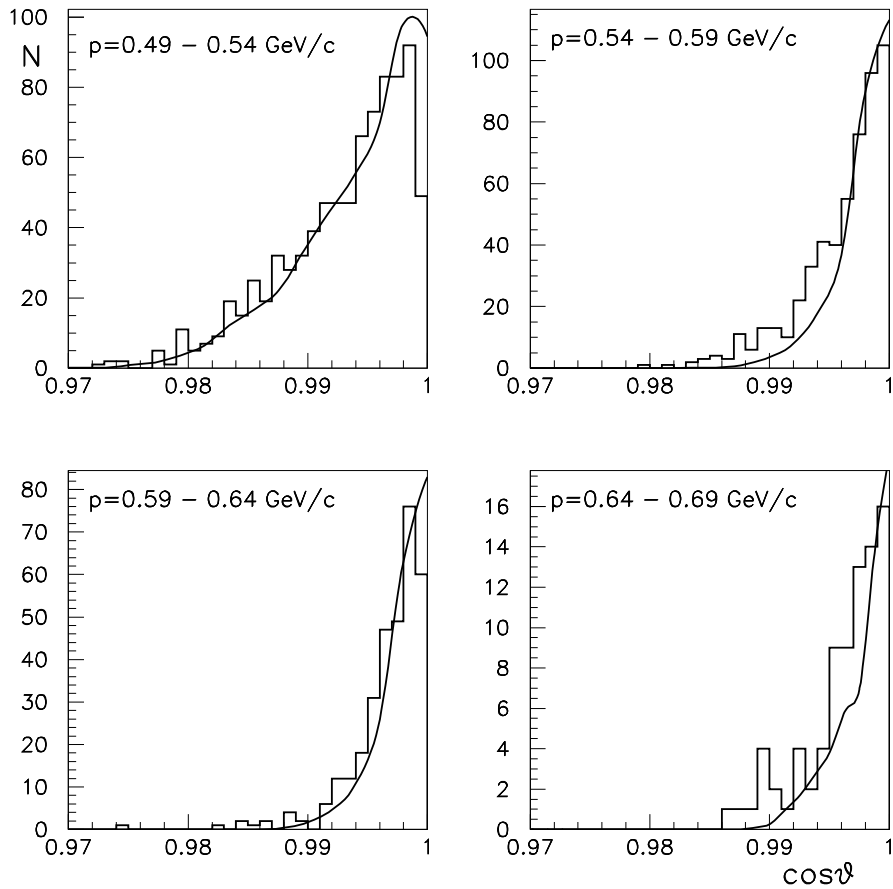}
\vspace*{-1.0cm}
\caption [] {Angular distributions of events in the
reaction (1) 
for various ranges of the outgoing positive pion momenta.
Solid curves represent the acceptance of the apparatus according to
the MC calculations assuming a uniform distribution of the cross section on
cos$\theta$ (arbitrary units).}

\label{diagram7}
\end{center}
\end{figure}

In  Fig.~\ref{diagram7} angular distributions of the DCX events taken with
the $S_{\pi}$ trigger at  $T_0=$ 0.59 GeV are shown in comparison with 
the Monte Carlo (MC) calculations normalized to the same number of events.
The calculations were performed assuming that the DCX differential 
cross section does not depend on a laboratory angle of $\pi^+$ emission. It
can be seen from  Fig.~\ref{diagram7} that within our angular acceptance and
statistics this assumption is in a good agreement with the data for all
intervals of $\pi^+$ momentum. It allows us to calculate the differential 
cross section as an average over the angular acceptance using the following 
expression
\begin{equation}
\frac{\mathrm{d^2}\sigma(T_0,\Delta T) }{\mathrm{d}\Omega \mathrm{d}T} =
\frac{N_{DCX}(T_0,\Delta T)} 
{(\rho l/A) N_{\mathrm{Av}} N_0 \Delta T \Delta \Omega(T_0,\Delta T)}
\frac{b}{k},\\
\end{equation} 
where $N_{DCX}(T_0,\Delta T)$ is the number of DCX candidates in the interval 
$\Delta T$ = 20 MeV; $A$ is the target mass number, $\rho$ and $l$ are the target 
density and thickness, $N_{\mathrm{Av}}$ is the Avogadro constant, and $N_0$ is the
beam flux.

The correction factor $b$ takes into account the proton and
the positron backgrounds.
In the case of the $S_{\pi}$ trigger, $b$ = 0.94$\pm$0.01,
0.87$\pm$0.04 and 0.80$\pm$0.09 at $T_0=$ 0.59, 0.75 and 1.1~GeV,
respectively. For the $S_{\pi}'$ trigger we have 0.71$\pm$0.06 and
0.71$\pm$0.09 at $T_0=$ 0.75 and 1.1~GeV.

\begin{figure}[bhtp]
\begin{center}

\includegraphics[bb=0 0 567 841,width=\textwidth]{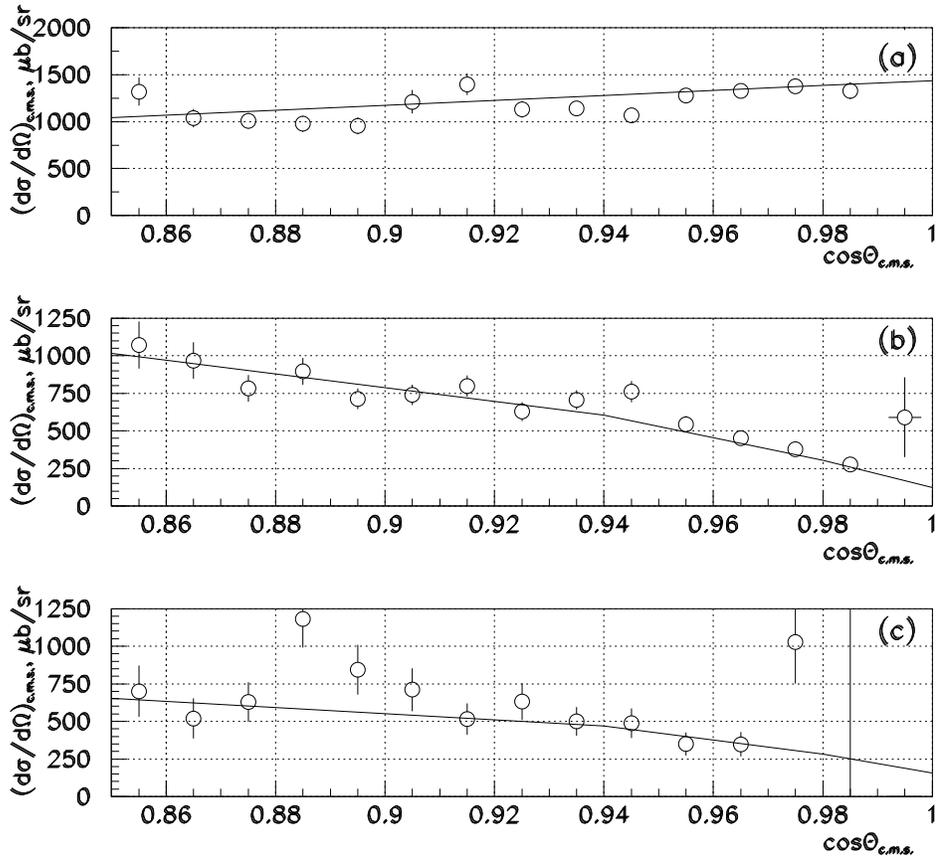}
\vspace*{-1.0cm}
\caption [] {The differential cross section for
the backward $\pi^{-}\mathrm{p}$ elastic scattering as
a function $\cos \theta_{c.m.s}$ ($\theta_{c.m.s}$ is a 
c.m.s. angle between the incoming pion and the outgoing proton)
at $T_{0}=$ 0.59 GeV (a), $T_{0}=$ 0.75 GeV (b) and
$T_{0}=$ 1.1 GeV (c). The solid lines are the PWA predictions taken 
from SAID (FA02 solution). The errors are statistical only.
}

\label{diagram7a}
\end{center}
\end{figure}

The  correction parameter 
$
  k = \prod\nolimits_{i} k_i, ~i=1,2,...,5
$
accounts for lepton contamination in the beam ($k_1$), beam halo on the target ($k_2$),
pion decay in the apparatus ($k_3$), pion absorption in the water target ($k_4$),
and pion absorption in the detector material ($k_5$). Depending on the beam 
momentum the value of $k$ varied from 0.53 to 0.75 with the uncertainty
$\pm$ 0.04;
$k_1$ = 0.81 -- 0.94 ($\pm$ 0.02),
$k_2$ = 0.77 -- 0.96 ($\pm$ 0.04), 
$k_3$ = 0.94 -- 0.97 ($\pm$ 0.03),
$k_4$ = 0.87 -- 0.92 ($\pm$ 0.02),
and $k_5$ = 0.94 -- 0.98 ($\pm$ 0.02).
The empty-target background was measured during each run and varied within 
3 $\pm$ 1 $\%$. To check the overall cross section normalization we applied 
a procedure analogous to formula (5) to obtain  the cross section of the 
backward elastic $\pi^-$p scattering with the proton trigger on a water target. 
The angular dependence
of this cross section is in an agreement with the partial-wave
analysis (PWA) of SAID \cite{arndt} shown in Fig.~\ref{diagram7a} (solid line is the
FA02 solution).

\begin{figure}[bhtp]
\begin{center}
\includegraphics[bb=0 0 567 841,width=\textwidth]{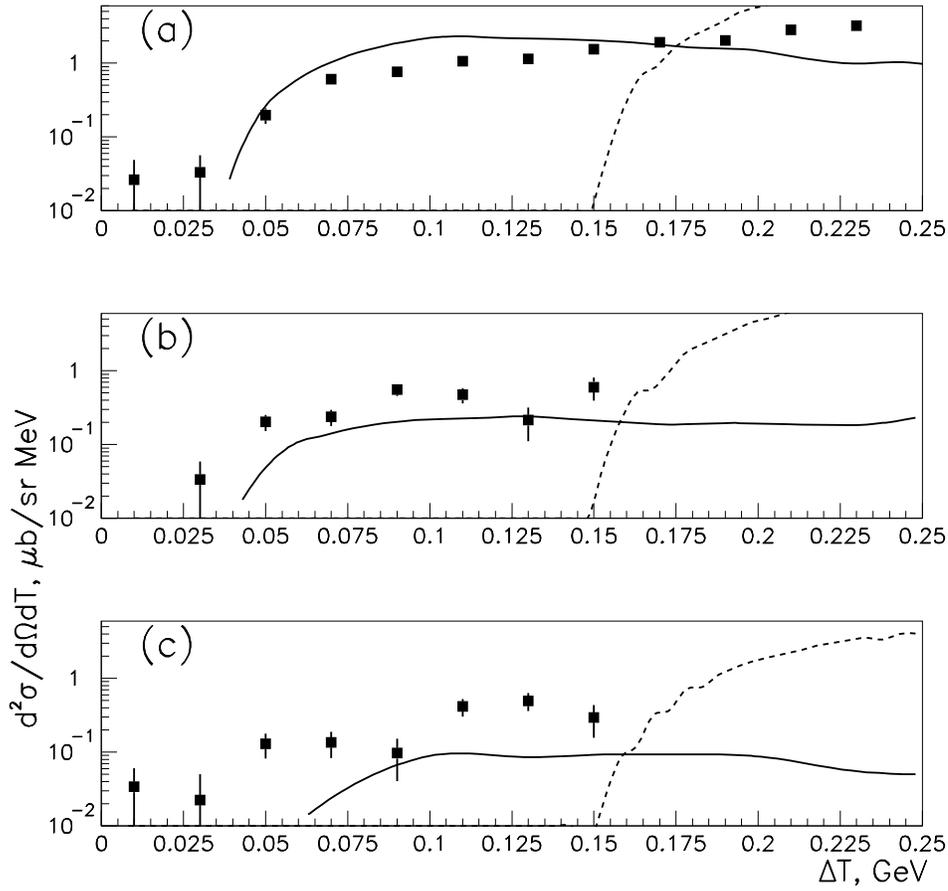}
\vspace*{-3.0cm}
\caption [] {
Differential cross section of the reaction
$\pi^{-16}$O$ \rightarrow \pi^+$X at $\langle \theta \rangle \approx 5^o$ as
a function of $\Delta T = T_0 - T$ for different pion beam
momenta : (a) $T_0$ = 0.59 GeV, measured with the
$S_{\pi}$ trigger; (b)  $T_0$ = 0.75 GeV, sum of both the
$S_{\pi}$ and the $S'_{\pi}$ triggers; (c) $T_0$ = 1.1 GeV,
sum of both the $S_{\pi}$ and the $S'_{\pi}$ triggers. The solid and dashed curves
were calculated in the framework of the MC cascade model for the sequential
mechanism (SSCX) and the additional pion production, respectively. 
}
\label{diagram8}
\end{center}
\end{figure}

The forward differential cross section of the reaction (1) at the beam
kinetic energies $T_0=$ 0.59, 
0.75 and 1.1 GeV was calculated according to formula (5) as the mean value 
of the cross sections on  H$_2$O and D$_2$O targets. The angular  
acceptance was $\theta  = $ 0 -- 14$^o$ for the $S_{\pi}$ 
trigger and  2 -- 10$^o$ for the 
$S_{\pi}'$ trigger with the mean value $\approx$5$^o$. At $T_0$ = 0.59 GeV
and with the $S_{\pi}$ trigger, the                                                                       
spectra were measured for two settings of the hodoscope $H_2$, 
target and  magnetic field, which provided small \cite{Yaf} and large  
$\Delta T$ acceptance with  approximately the same mean accepted angle.
In Fig.~\ref{diagram8}(a) the  $\Delta T$ distribution obtained 
for the DCX cross section with the
large acceptance is shown. The cross section grows monotonously in the 
range of $\Delta T$ from 0.03 to 0.25 GeV and has no peculiarity at the 
threshold  of the additional pion production process. Within our resolution 
we don't see a signal of the  $^{16}$C ground state \cite{ajzen} from the 
reaction $^{16}$O($\pi^-,\pi^+)^{16}$C(g.s.), which could appear at 
$\Delta T = $18.4 MeV. 

In Figs.~\ref{diagram8}(b) and (c), we show the summed spectra obtained with 
the $S_{\pi}$ and the $S_{\pi}'$ triggers at 0.75 and 1.1 GeV. The cross 
sections integrated over the  $\Delta T$ ranges from 0 to 80 MeV 
($\langle$d$\sigma$/d$\Omega \rangle_{80}$)\footnote{This $\Delta T$ 
interval was chosen to make the comparison with the inclusive data 
\cite{bur} at lower energy (LAMPF) which were kindly placed at our disposal 
by A.Williams.} and from 0 to 140 MeV  
($\langle$d$\sigma$/d$\Omega \rangle_{140}$)
with their statistical errors are  presented in Table 1 for the $S_{\pi}$
and the $S_{\pi}'$ 
triggers separately. As their values are in a good agreement for each beam 
energy we calculated the average values of the integrated cross sections; they 
are placed in the last line of the Table 1. The systematic errors are $\approx 10\%$.
The present values of $\langle$d$\sigma$/d$\Omega \rangle_{80}$ at 0.59 and
0.75~GeV are fully compatible with the preliminary results of
Ref.~\cite{Yaf}.
\begin{table}[bhpt]
\caption{The DCX cross sections integrated  over
the  $\Delta T$ range from 0 to 80 MeV,
$\langle$d$\sigma$/d$\Omega \rangle_{80}$, and over
the  $\Delta T$ range from 0 to 140 MeV,
$\langle$d$\sigma$/d$\Omega \rangle_{140}$,
obtained using pion  
triggers  $S_\pi$,
and  $S_\pi '$ and the averaged values of these cross sections (last line).}
\begin{center}
\begin{tabular}{l|ccc|ccc}
 & \multicolumn{3}{c|}{$\langle$d$\sigma$/d$\Omega \rangle_{80}$, $\mu
b/sr$} & \multicolumn{3}{c}{
$\langle$d$\sigma$/d$\Omega \rangle_{140}$, $\mu b/sr$}  \\[0.15cm]
\hline
$T_0$, GeV &0.59 &0.75 &1.1  &0.59 &0.75 &1.1 
\\[0.15cm]\hline
$S_{\pi}$
& 18.8$\pm$2.1   & 11.6$\pm$2.3  & 11.4$\pm$4.0
& 80.2$\pm$4.3   & 43.4$\pm$5.1  & 23.3$\pm$6.5  \\[0.25cm]
$S_{\pi} '$ &  &10.5$\pm$2.7 &6.4$\pm$1.9  &    &41.4$\pm$6.5 &31.4$\pm$4.7    
\\[0.15cm]\hline
$\langle S_{\pi}$+$S_{\pi} '\rangle$ &18.8$\pm$2.1    &11.1$\pm$1.8 &7.3$\pm$1.7 
&80.2$\pm$4.3    &42.6$\pm$4.0 &28.6$\pm$3.8\\[0.25cm] 
\end{tabular}
\end{center}
\end{table}

\section{Comparison with the SSCX model}
The first step towards an interpretation of the experimental results is
to compare them with a calculation based on the conventional SSCX mechanism,
represented by diagram (a) in Fig.~10 ($H_0=\pi^0$). 
For this purpose we have
used a Monte-Carlo cascade model constructed to describe pion induced multichannel
reactions (quasielastic, SCX, DCX, absorption and $\pi$ production) at pion
energies above 0.5~GeV~\cite{vacas}. This approach has been earlier applied
to the description of pion-nucleus inclusive reactions (including DCX) at
$T_\pi =85-350$~MeV~\cite{salcedo,vicente}. The basic idea is that a pion
propagating inside a nucleus can undergo elastic or quasielastic collisions,
be absorbed by two or three nucleons or produce another pion. The
probabilities per unit length of these processes are included in the model,
taking into account medium effects: Pauli blocking, Fermi motion and nucleon
binding energy. Furthermore, the renormalization of the $\pi$N
amplitude has been incorporated following Ref.~\cite{oset}. When compared to
the quasielastic data~\cite{vacas}, the model was able to reproduce the
quasielastic peak but underestimates pion production. This
discrepancy is common to other cascade codes like, for example the cascade
exciton model (CEM)~\cite{gudima} and has been addressed in a recent
publication \cite{mashnik}. However, the present
study, is mainly confined to the region where additional pion production is forbidden,
and which is well described by SCX.

The SSCX contribution to the cross section of the reaction (1) calculated 
in the framework of the MC cascade model \cite{vacas}
is shown in  Fig.~\ref{diagram8} by the solid curves.
Model predictions slightly overestimate the measured cross section for 
$\Delta T <$ 140 MeV at $T_0$ = 0.59 GeV and strongly underestimate it
at $T_0$ = 1.1 GeV.
As for the pion production
contribution, we can compare the theoretical predictions (see dashed 
curves in  Fig.~\ref{diagram8}) with the experimental results only at 
$T_0$ = 0.59 GeV, where we have a large $\Delta T$ acceptance. 
It can be seen that the calculation
overestimates the measured points. This discrepancy could be
related to the fact that the pion production mechanism, the scattering
angles and energies of the outgoing particles are assumed to follow the
3-body phase space distribution due to the lack of measured differential
cross sections~\cite{vacas}. 
 
Since there are  no other experimental data on the reaction (1) at our 
energies, we performed calculations for the charge symmetric reaction
\begin{equation}
\pi^+ + {}^{16}\mathrm{O} \rightarrow \pi^{-} + \mathrm{X} 
\end{equation} 
at $T_0=$ 0.4 -- 0.5 GeV for $\theta = 5^o$ to compare them with the 
available data taken at LAMPF \cite{bur}. In Fig.~\ref{diagram10} we see 
that the SSCX calculation (solid curve) overestimates these data. It is worth 
mentioning  
here, that the forward cross sections of the exclusive reactions  
$^{14}$C$(\pi^+,\pi^-)^{14}$O and $^{18}$O$(\pi^+,\pi^-)^{18}$Ne, 
which were calculated within Glauber theory without free parameters 
\cite{oset}, also appeared to be higher than the measured ones at
$T_0=$ 300 -- 525 MeV  \cite{williams}. The authors of \cite{oset}
succeeded in reaching an agreement with the data of  \cite{williams}
after they took into consideration a medium polarization, which led to
the renormalization of the charge exchange $\pi \mathrm{N}$ amplitude. Taking
this effect into account in our model substantially decreases the value
of the DCX cross section for the reaction on $^{16}$O (see dashed curves
in Fig.~\ref{diagram10}) 
and improves an agreement
with the data at $T_0 =0.59$~GeV. However for higher energies, as it will be
shown in the next Section, it results in larger deviation from the data.

\section{Energy dependence}
In Fig.~\ref{diagram11} and Fig.~\ref{diagram12},
the energy dependence 
of the partially integrated pion DCX cross sections 
$\langle$d$\sigma$/d$\Omega \rangle_{80}$ 
and $\langle$d$\sigma$/d$\Omega \rangle_{140}$ is shown. The 
results from the last line of Table 1 are presented by the black circles,
while the experimental data of \cite{wood} for the reaction (1) at 
$T_0 = $ 0.18, 0.21 and 0.24 GeV and $\theta = 25^o$, and the results of
\cite{bur} on the reaction (6) are shown by empty circles and squares, 
respectively. The DCX cross section decreases with
energy by about a factor of six 
from 0.18 to 1.1 GeV, but the cross section calculated within the SSCX model for 
$\theta = 5^o$ (solid curves in Fig.~\ref{diagram11} and Fig.~\ref{diagram12})
falls considerably faster in the region of 0.6 -- 1.1 GeV. This behaviour 
reflects the fast decrease of a single charge exchange  $\pi \mathrm{N}$ 
amplitude in this energy region. At 0.75 and 1.1 GeV the measured cross 
section is significantly larger than expected from the conventional SSCX 
mechanism. Therefore, other (nonconventional) approaches to the high energy DCX 
or a substantial modification of 
$\pi \mathrm{N}$ amplitudes
are needed to explain the observed discrepancy between theory and experiment. 

The results of the theoretical calculations within the SSCX mechanism with medium 
polarization are shown by dashed-dotted curves marked with stars.  
The shape of these curves with and without  medium polarization 
effect is similar and the  discrepancy of the theory and the experiment gets 
even larger when it is taken into account.

Finally, let us note that the cross section of the exclusive reaction 
$^{16}$O$(\pi^+,\pi^-)^{16}$Ne, that is a small part of the inclusive 
process (6), falls in the range 0.18 -- 0.25 GeV \cite{gilman}-\cite{beatty1},
then increases up to 0.3 GeV and stays constant within experimental errors up to 
0.5 GeV  \cite{beatty2}.

\section{Other approaches to pion DCX}
\begin{figure}[bhtp]
\begin{center}
\includegraphics[bb=0 0 567 841,width=\textwidth]{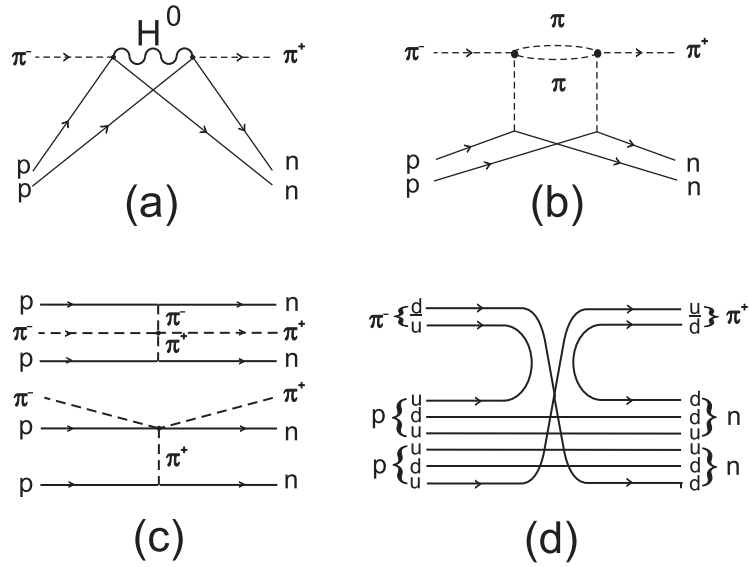}
\vspace*{-3.0cm}
\caption [] {Diagrams contributing to pion double charge exchange
on a nucleus --\\
(a) sequential charge exchanges on two different protons:\\
$H^0 = \pi^0$, sequential single charge exchanges (SSCX),\\
$H^0 = \eta^0, \rho^0...~$ quasielastic rescatterings,\\
$H^0 = n\pi$, inelastic Glauber rescatterings ($H^0 = 2\pi^0, \pi^+\pi^-,...)$,\\
(b) inelastic rescatterings with $H^0 = 2\pi$ in the OPE model,\\
(c) meson exchange currents (MEC),\\
(d) short-range $NN$ correlations.} 

\label{diagram9}
\end{center}
\end{figure}

\begin{figure}[bhtp]
\begin{center}
\includegraphics[bb=0 0 567 841,width=\textwidth]{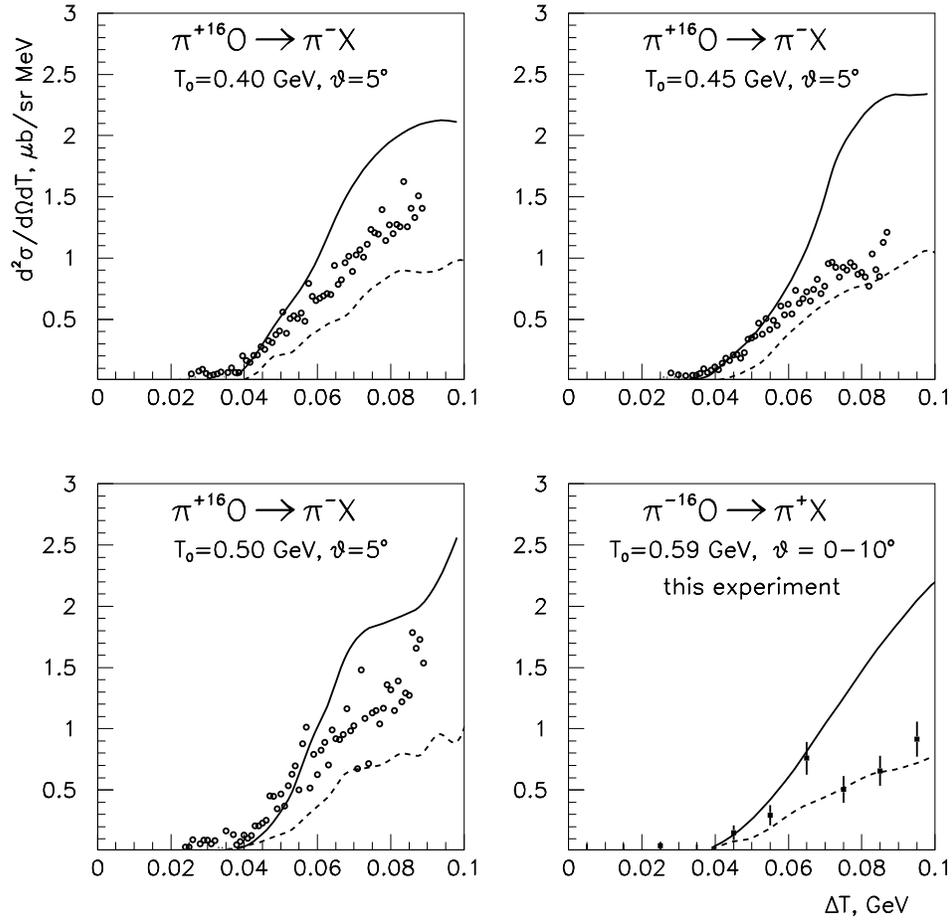}
\vspace*{-1.0cm}
\caption [] {Double-differential cross sections for  
$\pi^{+16}\mathrm{O} \rightarrow \pi^-$X \cite{bur}
and $\pi^{-16}\mathrm{O} \rightarrow \pi^+$X
reactions versus $\Delta T$. 
Curves are calculated in the framework
of SSCX mechanism with (dashed curve) and without (solid curve)
medium polarization. 
}
\label{diagram10}
\end{center}
\end{figure}

\begin{figure}[bhtp]
\begin{center}
\includegraphics[bb=0 0 567 841,width=\textwidth]{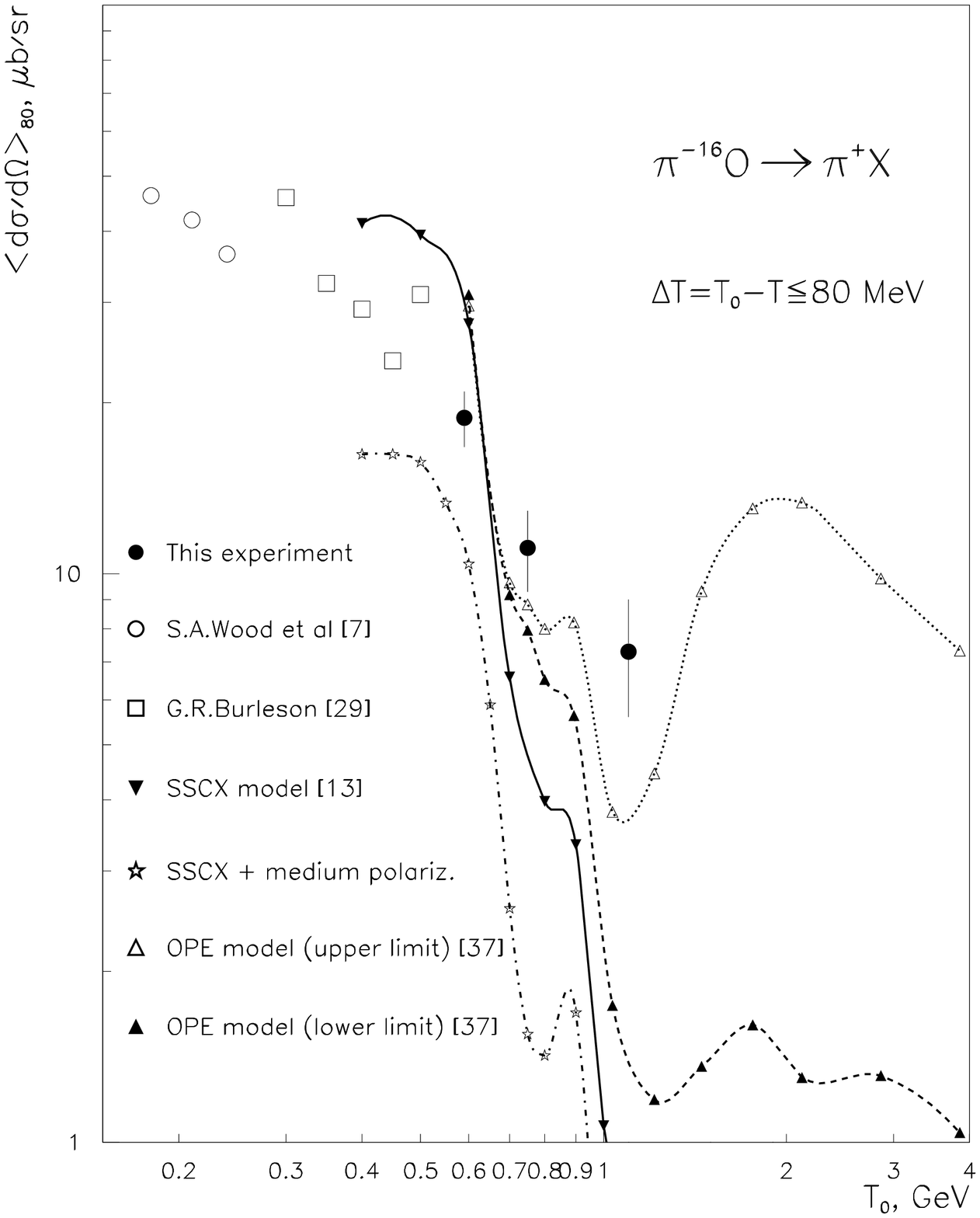}
\vspace*{-3.0cm}
\caption [] {Energy dependence of the DCX cross section integrated over
the $\Delta T$ range from 0 to 80 MeV.
} 
\label{diagram11}
\end{center}
\end{figure}

\begin{figure}[bhtp]
\begin{center}
\includegraphics[bb=0 0 567 841,width=\textwidth]{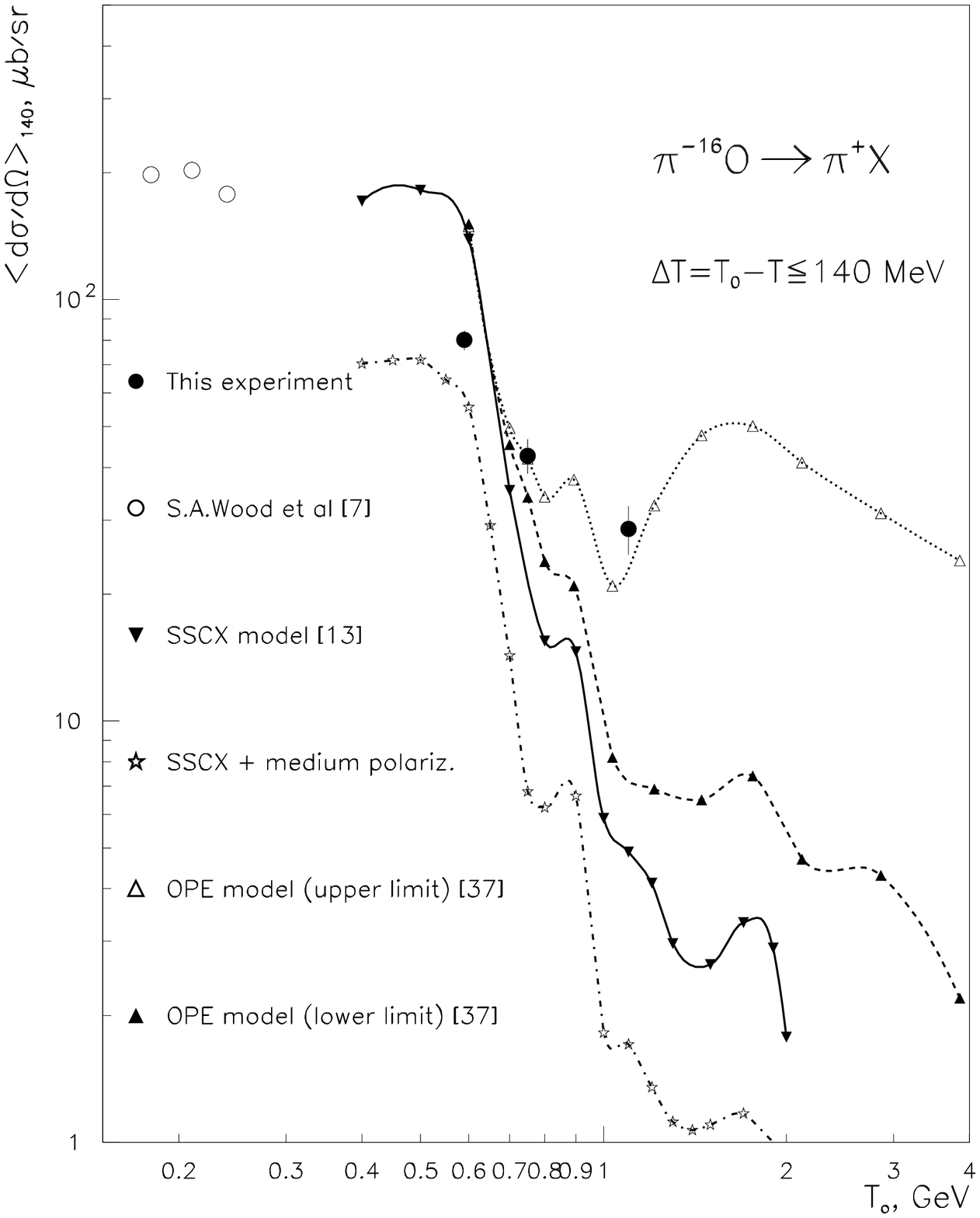}
\vspace*{-3.0cm}
\caption [] {Energy dependence of the DCX cross section integrated over
the $\Delta T$ range from 0 to 140 MeV.
} 
\label{diagram12}
\end{center}
\end{figure}

We observed a large deviation of the measured DCX cross section from the SSCX 
prediction at high energy. 
Two new approaches  have been tried to explain the phenomenon.
The first 
calculations of the contribution of 
meson exchange 
currents (MEC) to DCX (see diagrams (c) in Fig.~\ref{diagram9}) in this 
energy region have been performed in \cite{luis} for the exclusive reaction 
$^{18}$O$(\pi^+,\pi^-)^{18}$Ne. At $T_0 >$ 0.5 GeV they showed a weak energy
dependence comparable with the one seen in our experiment for inclusive DCX.
This reflects the very weak energy dependence of the MEC amplitudes.
However, the absolute values of MEC amplitude appeared to be too 
small to be relevant.

A new idea was proposed  in Ref. \cite{kaid}. It was shown that the Glauber
inelastic rescatterings (IR) (see diagram (a) in Fig.~\ref{diagram9} with
$H^0$ being a multipion state) gave an important contribution to 
the inclusive DCX cross section at energies $T_0 \ga$ 0.6 GeV and allowed 
to understand the relatively slow decrease of the cross section with energy.
The cross section of the reaction (1) was expressed as the sum of two terms,
one with an intermediate $\pi^0$ (SSCX) and another with an intermediate
$2\pi$ state (IR contribution). The IR part  was calculated \cite{kakru} 
in the framework of the Gribov-Glauber approach to DCX 
 using the experimental
data on the reactions $\pi^- \mathrm{p} \to \pi \pi $N \cite{brody} and
the one pion exchange
(OPE) model (see diagram (b) in Fig.~\ref{diagram9}).
The
dotted and dashed curves in Figs.~\ref{diagram11} and \ref{diagram12}
correspond to upper and lower boundaries of the theoretical estimation. The
upper limit is close to the experimental data, especially for
$\langle$d$\sigma$/d$\Omega \rangle_{140}$, which represents a considerable
improvement with respect to the SSCX calculation. 
Although at present the uncertainty in these predictions is rather large the 
IR contribution seems to be able to explain the observed energy dependence. 

The analysis of \cite{kakru} indicates that at higher energies the
low-boundary curve (the dashed one) is more justified and that in this case 
the average distances, $d$, 
between nucleons participating in the interaction are relatively small: 
$d \sim (2m_{\mathrm{N}}\Delta T)^{-1/2}$. These nucleons are rather closely
correlated and can be even considered as a 6-quark system corresponding 
to the diagram (d) in Fig.~\ref{diagram9}.

Although the diagrams (b) and the upper one in (c) 
(see Fig.~\ref{diagram9}),  which stand  
for IR in the OPE model and MEC respectively, look similar, we would like to stress 
their principal difference: in the OPE model we treat the 
$\pi^-\pi^+$-scattering amplitude as a function of $M^2$ ($M$ is the mass 
of the intermediate state $H^0$) and integrate it over this variable while 
in the MEC model the $\pi^-\pi^+$ amplitude is approximated as a point-like 
interaction taken at the threshold ($M = 2m_{\pi}$). Due to the soft pion 
theorems the last amplitude is small and thus leads to small modifications 
of SSCX predictions \cite{luis}. We take into account both real and 
imaginary parts of $\pi^-\pi^+$ amplitude in the regions of $M^2$ where 
they are not small, so it is not surprising that our calculations for DCX cross 
sections are substantially higher than SSCX predictions (even for the dashed 
curves in Fig.~\ref{diagram11} and Fig.~\ref{diagram12}).

\section{Conclusion and outlook}

In summary, we have performed the first measurement of the cross
section for the forward inclusive reaction
$\pi^- +^{16}\mathrm{O} \rightarrow \pi^{+} +$ {X} at energies $T_0  = $
0.59 -- 1.1 GeV, in the region where additional pion production is
kinematically forbidden. We have found that the cross section decreases with
energy considerably slower than it is predicted by the conventional DCX
mechanism of two (or more) sequential single charge exchanges 
with a $\pi^0$ in the intermediate state. At $T_0 =
1.1$ GeV, the experimental cross section is a factor of five larger than the
theoretical one. This discrepancy implies that new mechanisms should be
invoked in the region of $T_0 \sim 1$ GeV. 
The contribution of two-pion
intermediate states, obtained within the Glauber-Gribov framework, seems to
play an important role
but the large 
theoretical uncertainty of the
calculation does not allow to come to more definite conclusion. 
Experimental data at higher energies will be extremely helpful to clarify the
situation and to constrain this model.

\begin{ack}
We are grateful to the staff of the 3m spectrometer, of the ITEP
PS as well as the PSP-2 group for the assistance during the 
experiment. We are pleased to thank G.A.Leksin for his interest to this
study and for the discussions of the results.
A.P.K. thanks 
I.S. and I.I.Tsukerman for 
permanent 
support. This work 
was supported in part by RFBR Grants 98-02-17179, 00-15-96545,
00-15-96786 and 01-02-17383,
and by Grant INTAS-93-3455.
\end{ack}


\clearpage

\clearpage

\clearpage

\clearpage


\clearpage


\clearpage


\clearpage

\clearpage

\clearpage

\clearpage

%

\clearpage

\clearpage


\end{document}